\documentclass[11pt,twoside]{article}
\usepackage{./asp2014}

\aspSuppressVolSlug
\resetcounters

\begin{document}

\title{Stripped red giants - Helium core white dwarf progenitors and their sdB siblings}
\author{Ulrich Heber$^1$ 
\affil{$^1$ Dr. Karl Remeis-Sternwarte \& ECAP, Astronomical Institute, University of Erlangen-N\"urnberg, Bamberg, Germany \email{heber@sternwarte.uni-erlangen.de}}}

\paperauthor{U. Heber}{heber@sternwarte.uni-erlangen.de}{}{University of Erlangen-N\"urnberg}{Dr. Karl Remeis-Sternwarte \& ECAP, Astronomical Institute}{Bamberg}{}{D 96049}{Germany}
\begin{abstract}
Some gaps in the mosaic of binary star evolution have recently been filled by the discoveries of helium-core white dwarf progenitors (often called extremely low mass (ELM) white dwarfs) as stripped cores of first-giant branch objects. Two varieties can be distinguished. One class is made up by SB1 binaries, companions being white dwarfs as well, another class, the so-called EL CVn stars, are composite spectrum binaries, with A-Type companions. Pulsating stars are found among both classes. A riddle is posed by the apparently single objects. There is a one-to-one correspondence of the phenomena found for these new classes of star to those observed for sdB stars. In fact, standard evolutionary scenarios explain the origin of sdB stars as red giants that have been stripped close to the tip of first red giant branch. A subgroup of subluminous B stars can also be identified as stripped helium-cores of red giants. They form an extension of the ELM sequence to higher temperatures. Hence 
low mass white dwarfs of helium cores and sdB stars in binaries are close relatives in terms of stellar evolution.  
\end{abstract}

\section{Introduction}

The mass distribution of DA white dwarfs displays two small populations in addition to the common C/O white dwarfs 
with a sharply peaked mean mass of 0.649 M$_\odot$ \citep{2015MNRAS.446.4078K}.  
The existence of low mass white dwarfs, that is  
with masses below the canonical mass for helium ignition (0.47 M$_\odot$),
was suggested by \citet{1992ApJ...394..228B}. Those white dwarfs were, therefore, considered as helium core white dwarfs.
Later studies of larger samples \citep{2005ApJS..156...47L} confirmed the existence of a population of white dwarfs with a low mean mass of $\approx$0.4~M$_\odot$ (henceforth LMWD). 
Since the evolutionary time for a single star of such a low mass to evolve into a helium-core white dwarf would exceed the age of the Universe by far, \citet{1992ApJ...394..228B} concluded that they must be in close binaries, which was, indeed, confirmed {{in}} many cases by radial velocity studies. In this scenario the helium-core was exposed by stripping the envelope via mass transfer to a companion. 
However, not all low-mass
WDs are found in binary systems \citep{2000MNRAS.319..305M,2007ASPC..372..387N}.
\citet{2011ApJ...730...67B} estimate that the fraction of apparently single 
low-mass white dwarfs is less than 30\%. The origin of these stars remains a puzzle.

It is premature to assume that white dwarfs with masses below the canonical mass for core helium burning are of helium composition, because intermediate mass stars ($\geq$ 2.3 M$_\odot$) may ignite helium at the tip of the RGB in non-degenerate conditions at core masses as low as 0.33 M$_\odot$ \citep[e.g.][]{2008A&A...490..243H}. 
Hence, a bona-fide helium-core object has to be less massive than 0.33 M$_\odot$.

Surveys for white dwarfs were tailored for high gravity stars and even in the largest white dwarf sample \citep[from the SDSS survey,][]{2015MNRAS.446.4078K} 
does not contain white dwarfs with masses below 0.3 M$_\odot$. Recently, a dedicated survey \citep[the ELM survey,][and references therein]{2013ApJ...769...66B} discovered several dozen objects of very low mass ($\approx$0.15 to 0.3 M$_\odot$) by exploring a sparsely populated region in a two-color diagram . 
White dwarfs are usually considered to have surface gravities exceeding $\log$ g = 7 resulting in strongly Stark broadened spectral lines. The newly discovered stars have considerably lower gravities ($\log$ g = 4.5 to 7) and, therefore, narrower lines, quite different from classical white dwarfs.

Hot subluminous stars are extreme horizontal branch stars, thus burning helium in their cores. Many of them occur in close binaries, which suggest that they are also stripped cores of RGB stars, similar to binary LMWDs and ELM WDs. The difference in evolution is just, that the progenitor of the sdB stars has climbed the RGB to the tip in order to ignite 
helium in the core, while LMWDs and ELM WDs depart from the RGB earlier on (see Fig. \ref{fig1}).


\articlefigure[angle=270,width=0.62\textwidth]{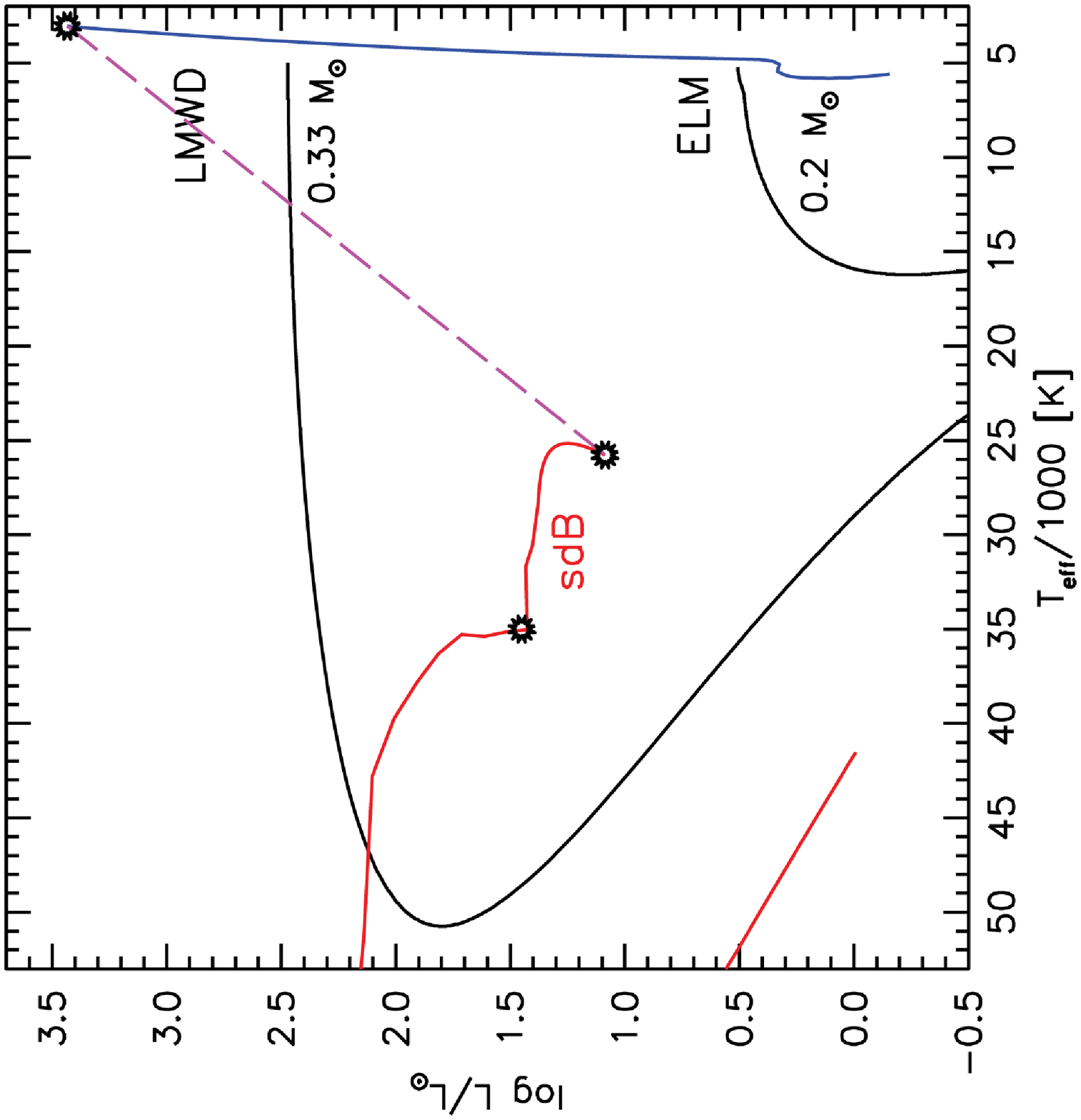}{fig1}{Sketch of the evolution of an 1 M$_\odot$ star (blue) from the zero age main sequence to the tip of the red giant branch, stripped at a core mass of $\approx$ 0.2M$_\odot$ to form an extremely low mass (ELM) white dwarf; stripped at $\approx$ 0.33 M$_\odot$ to form a low mass white dwarf (LMWD), and at the onset of the helium flash to form an sdB star (dashed line). The beginning and the end of the core helium burning of the sdB evolution (red line) is also marked by asterisks. From \citet{2016PASP..128h2001H}.}

\section{ELM white dwarfs: Low mass white dwarfs below the helium burning limit}\label{sect:elm}

In fact ELM WDs were discovered well before the ELM survey emerged, namely as companions to millisecond pulsars.
\citet{1996ApJ...467L..89V} 
showed that the companion of a neutron star in the millisecond pulsar {PSR J1012+5307} is a low-mass helium-core object, for which \citet{1998A&A...339..123D} determined a mass of M=0.19$\pm$0.02 M$_\odot$ and a cooling age of 6$\pm$1 {Gyr}. 
\citet{2014A&A...571L...3I} list another eight millisecond pulsars in short period binaries with a low-mass ($<$0.21 M$_\odot$) proto-helium white dwarf.

Because of the their low masses they must be core helium objects formed in a close binary via envelope stripping. However, their low gravities indicted that the stars are bloated and many may not have reached the cooling sequence yet. This is because such a star retains an envelope ($\approx$ 0.01 M$_\odot$) thick enough to sustains hydrogen burning and therefore the transition times to the cooling sequence are long   
\citep[(up to several billion years, depending on stellar mass, see e.g.][]{2014A&A...571L...3I}.

Different names were coined to describe these low-mass objects: ``helium core white dwarf progenitors'' \citep{2003A&A...411L.477H}, 
``proto-helium white dwarfs'' \citep{2014A&A...571L...3I},``thermally bloated, hot white dwarf" \citep{2011ApJ...728..139C}, 
``pre-He-WD'' \citep{2013Natur.498..463M}, 
or ``Extremely low mass white dwarfs (ELM)'' \citep{2010ApJ...723.1072B}, 
which can be regarded as synonymous. The term ELM white dwarf apparently has won recognition. 
Because surface gravities of the class members differ enormously, it is useful to divide the class into pre-ELM and ELM white dwarfs using the turn-over point of their evolution in the T$_{\rm eff}$ - $\log$ g diagram (see Fig. \ref{fig2}) as reference, with pre-ELM WDs having lower and ELM WDs higher gravities

It was expected that these stars form in close binaries, and, therefore, they were monitored for radial-velocity variations with a very high success rate. Most were found to have orbital periods shorter than one day. Their companions were identified to be white dwarfs as well.

\section{Subluminous B stars}

Most of the B-type subdwarfs were identified as helium burning stars of about half a solar mass
at the blue end of the horizontal branch, the so-called Extreme Horizontal Branch \citep[see ][]{2009ARA&A..47..211H,2016PASP..128h2001H}. 
 Binary evolution through mass transfer and 
common-envelope ejection must be important for sdB stars because of the
the fraction of close binaries with periods of less than ten days is as high as 50\%. Single sdB stars can be formed via a merger of two helium white dwarfs
\citep{2003MNRAS.341..669H,2016MNRAS.463.2756H} as backed up by recent observations. \citet{2011ApJ...735L..30P} namely discovered CSS 41177 to be a detached eclipsing double white dwarf binary in a 2.78 hr period orbit. Both members are helium core white dwarfs as suggested by their masses of M$_1$ = 0.283 M$_\odot$ and M$_2$ = 0.274 M$_\odot$). The system will be driven into a merger by gravitational wave radiation merge and will form a single sdB star in about 1.1 Gyrs.

\section{Proto-helium white dwarfs imitating sdB stars}

The transition track of a proto-helium white dwarf from the red giant branch to the cooling sequence leads through a wide area of the (T$_{\rm eff}, \log~g)$ plane and across the main sequence and the (extreme) horizontal branch (see Fig. \ref{fig2}). Therefore, proto-helium white dwarfs may be confused with normal stars. In fact a few such cases are known\footnote{The most extreme case, HZ~22, even mimics an early type main-sequence star.}, in particular the sdB stars HD~188112, KIC 06614501, SDSS J0815+2309,
SDSS~J1625+3632 and GALEX J0805$-$1058 imitate sdB stars (see Fig. \ref{fig2}). The sdB+WD binary HD~188112 is a well-studied example. 

\articlefigure[width=0.6\textwidth]{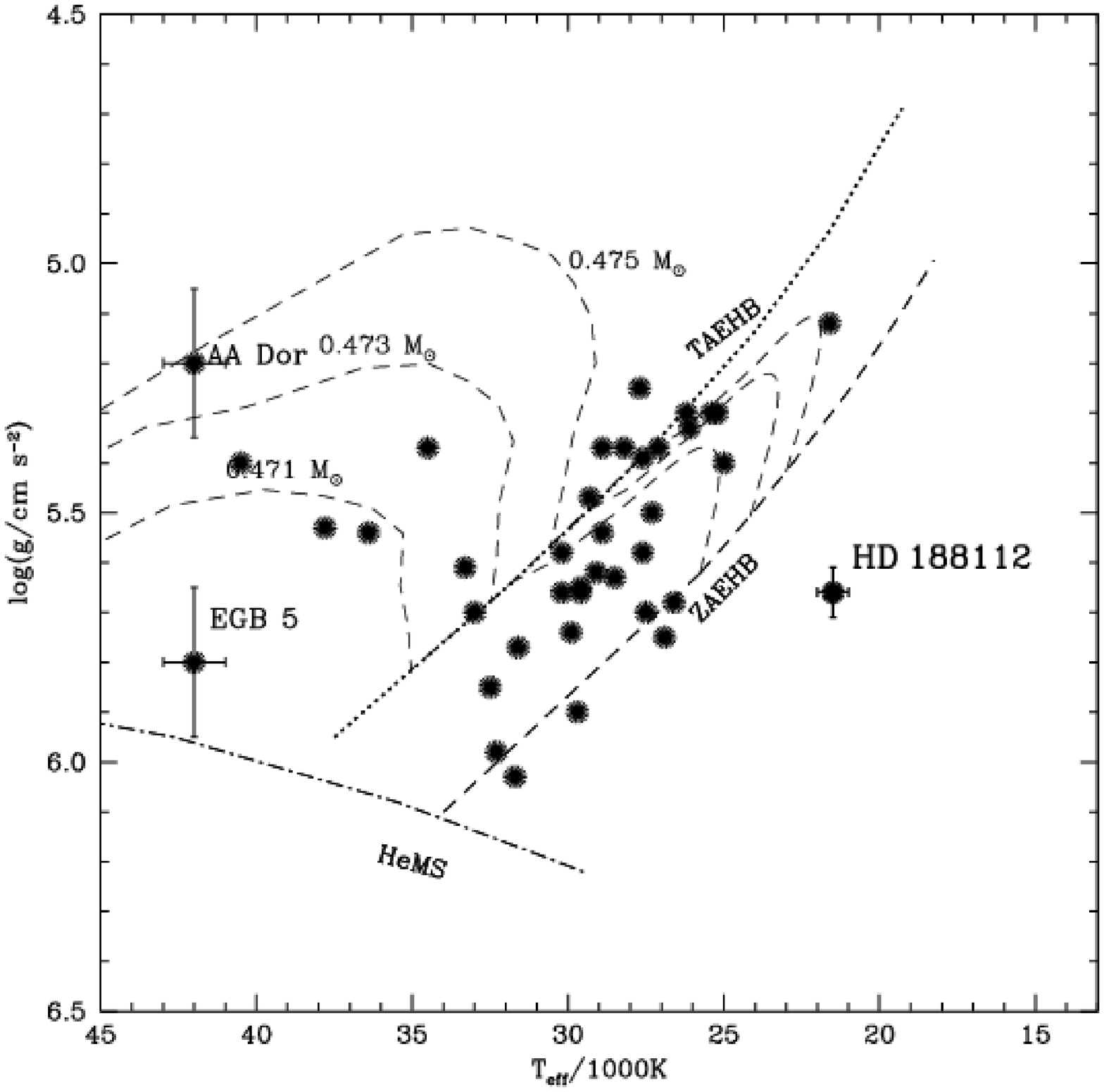}{fig3}{Position of HD 188112 in the ( $ {T_{\rm eff}}$, log g) plane and comparison to sdB stars in close binary systems with known periods and evolutionary models for post-EHB evolution (from \citep{2003A&A...411L.477H}.} 

\subsection{HD~188112 - a (pre-) ELM WD}

\citet{2003A&A...411L.477H} found the sdB star HD~188112 to be of somewhat higher gravity than a core-helium burning EHB star (see Fig. \ref{fig3}) from a quantitative spectral analysis. Its Hipparcos parallax  combined with the spectroscopic gravity allowed the mass to be determined to be ~0.23 M$_\odot$, consistent with evolutionary model predictions for stripped RGB stars of such a mass.
Hence, it was concluded that HD~188112 is a helium core white dwarf progenitor, i.e. an ELM white dwarf close to the turn-over point of the corresponding evolutionary model. Whether the stars should be classified as an ELM or pre-ELM depends on the evolutionary tracks used. Meanwhile
the star has been studied in great detail because it is by far brighter than any other ELM WD known. 
\citet{2016A&A...585A.115L} measured the rotational broadening of metallic lines in our UV spectra and derived vrot sin i = 7.9 $\pm 0.3$ km s$^{-1}$. Assuming tidally locked rotation, the companion masses could be constrained to lie between 0.92 and 1.33 M$_\odot$. Recent statistical investigations by \citet{2014ApJ...797L..32A} and \citet{2015A&A...575L..13B} on the ELM companions' mass distribution found a mean mass of ~0.75 $\pm$ 0.25 M$_\odot$ 
The companion of HD~188112 would lie near the high-mass edge for the ELMs' distribution. Also a neutron star companion can not be excluded, as it would require an inclination angle only slightly lower than required for tidal locking. The star would then be rotating somewhat faster than synchronized. Thus, a search for X-ray emission would be worthwhile.

It is interesting to note that HD~188112 differs from typical sdB star in its  metal content which is 
significantly lower than that of the latter \citep{2016A&A...585A.115L}. Extremely low-mass white dwarfs also show metals in their spectra, e.g. all the ELM WDs with log g $<$ 6.0 show Ca II K lines \citep{2014ApJ...794...35G}, Detailed abundance
analyses, however, are available for only a handful of objects with a diversity of abundance pattern, as in the case of sdB stars. HD~188112, again stands out, because abundance for no less than 15 chemical elements from Mg to Pb have been determined with high precision and accuracy. With respect to Mg and Ca HD~188112 is found to lie at the low end of the ELM distribution 
of \citet{2014ApJ...794...35G}.
  
\articlefigure[width=0.68\textwidth]{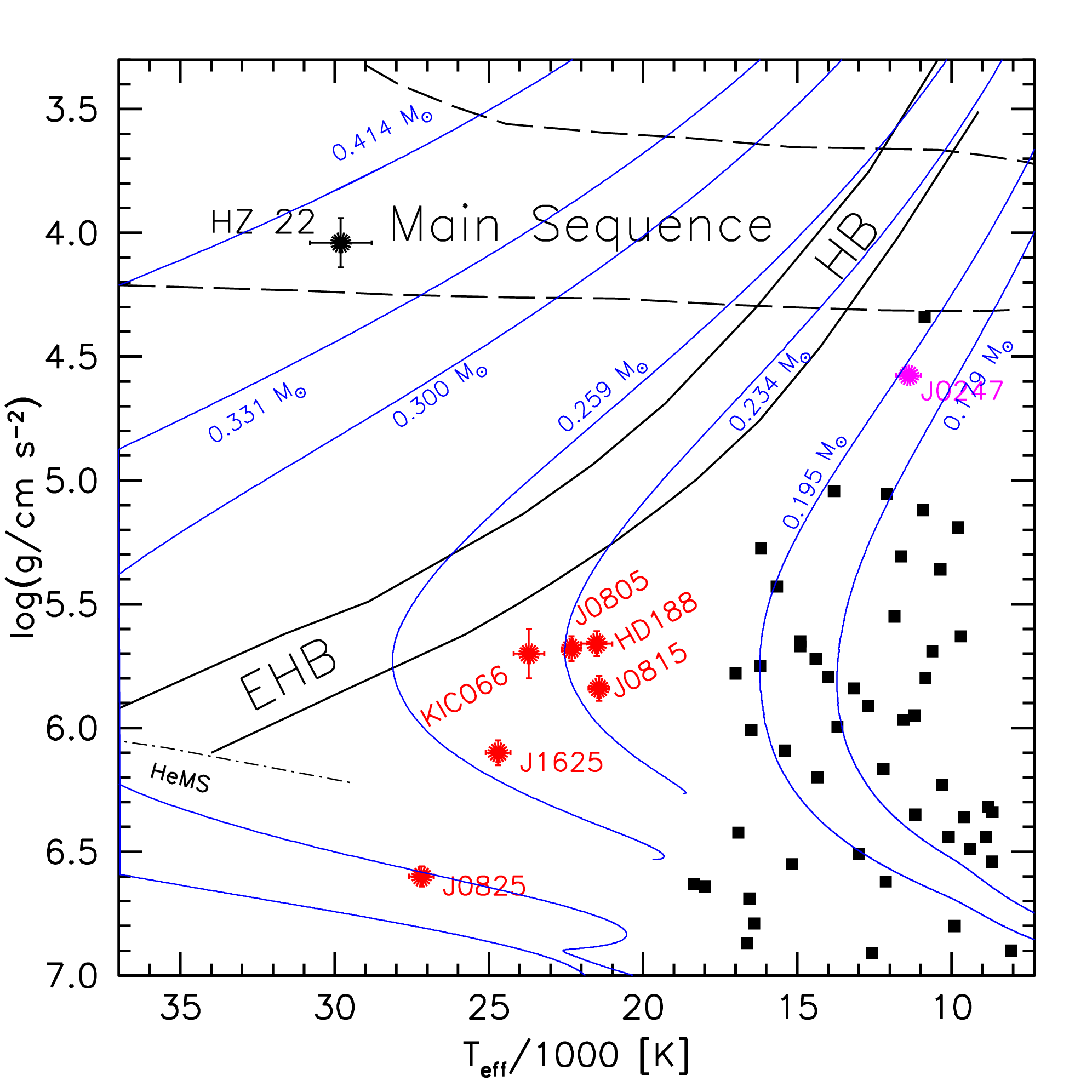}{fig2}{Distribution of (pre-) ELM white dwarfs in the 
T$_{\rm eff}$ -- $\log$ g - diagram. The horizontal branch (HB) and the extreme horizontal branch (EHB) are bracketed by the full drawn lines. The dashed lines depict the main-sequence band and the dashed-dotted line marks the helium main sequence. Tracks for the evolution of proto-helium white dwarfs are from 
\citet{1998A&A...339..123D} and labelled {{with}} their mass. 
 Objects are named pre-ELM if their gravity is lower than that at the turn-over of the corresponding evolutionary track (highest temperature), close to $\log$~g=6.0. Filled squares are (pre-) ELM WDs from the ELM survey. 
(Pre-) ELM WDs that spectroscopically imitate sdB stars are highlighted in red (below the EHB). Note that they lie close to the turn-over points of the evolutionary tracks.  From \citet{2016PASP..128h2001H}.
 }

\section{Pre-ELM WDs with main sequence companions: The EL~CVn stars}

The ELM survey revealed the existence of binaries consisting of ELM-WDs with degenerate companions (white dwarfs
or neutron stars as pulsars). Recently, a new class of binaries was discovered, named EL CVn stars after its prototype \citep{2011MNRAS.418.1156M}, which consists of pre-ELM WDs with A-and F-type companions. The number of class members has grown to 17 by now \citep{2014MNRAS.437.1681M}.
Again a similarity to the sdB class may be noticed, because a considerable fraction of sdB stars have dwarf F/G/K-type companions.


\section{Close relatives: Subluminous B stars, LMWDs and (pre-)ELM white dwarfs}

As discussed above some subluminous B stars may actually be helium-core white dwarfs of the ELM type or in an earlier phase of evolution (pre-ELM). In addition, the class of (pre-) ELM WDs shares a lot of phenomena also observed in sdB stars, e.g. diffusion dominated metal abundance pattern. Multi-periodic pulsations have recently been discovered both in ELM \citep{2012ApJ...750L..28H,2013ApJ...765..102H} and pre-ELM WDs \citep{2013Natur.498..463M,2016ApJ...822L..27G}, while two classes of sdB pulsators have been found previously
\citep{1997MNRAS.285..640K,2003ApJ...583L..31G}. The analysis of the pulsation spectra paves the way to asteroseismology, that is to probe the internal structure of these stars and derive total as well as envelope masses. Mergers driven by gravitational wave radiations may lead to single objects. The further evolution of these systems relates i.a. to the formation of AM CVn binaries and possibly to type Ia supernova explosions.  A more detailed discussion can be found in \citet{2016PASP..128h2001H}.



\end{document}